\documentclass[11pt]{article}
\usepackage{amsmath,epsf,amssymb,latexsym,enumerate,cite,shadow,array,mcite,units,verbatim,amsmath}
\usepackage{slashed}
\usepackage{graphicx,wasysym}

\DeclareFontFamily{U}{rsf}{} \DeclareFontShape{U}{rsf}{m}{n}{
  <5> <6> rsfs5 <7> <8> <9> rsfs7 <10-> rsfs10}{}
\DeclareMathAlphabet\Scr{U}{rsf}{m}{n} \makeatletter
\@addtoreset{equation}{section} \makeatother

\def\be{\begin{equation}}
\def\ee{\end{equation}}
\def\ba{\begin{array}}
\def\ea{\end{array}}

\newcommand{\bea}{\begin{eqnarray}}
\newcommand{\eea}{\end{eqnarray}}

\def\K{K{\"a}hler}
\def\bi{{\bar I}}

\usepackage{ifpdf}
\ifx\pdfoutput\undefined
   \pdffalse
\else
   \pdfoutput=1
   \pdftrue
  \usepackage[pdftex]{hyperref}
\def\u0{{\underline 0}}

\def\url{{\underline {r+\ell}}}

\usepackage{color}
\usepackage[makeroom]{cancel}
\usepackage{graphicx}
\usepackage{amssymb,amsthm}
\usepackage{hyperref}

\def\bi{\begin{itemize}}
\def\ei{\end{itemize}}
\def\be{\begin{equation}}
\def\ee{\end{equation}}

\begin{document}

\begin{titlepage}

\hskip 1cm

\vskip 2cm

\begin{center}

{\huge \bf{Inflation and Dark Energy \\ \vskip 0.3 cm with a Single Superfield}}

\

{\bf  Andrei Linde},$^{1}$\  {\bf Diederik Roest}$^{2}$ and {\bf Marco Scalisi}$^{1,2}$   
\vskip 0.5cm
{\small\sl\noindent $^{1}$SITP and Department of Physics, Stanford University, Stanford, CA
94305 USA\\
$^{2}$Van Swinderen Institute for Particle Physics and Gravity, University of Groningen,\\
Nijenborgh 4, 9747 AG Groningen, The Netherlands\\
}
\end{center}
\vskip 1 cm

\begin{abstract}
We discuss the possibility to construct supergravity models with a single superfield describing inflation as well as the tiny cosmological constant $V \sim 10^{{-120}}$. One could expect that the simplest way to do it is to study models with a supersymmetric Minkowski vacuum and then slightly uplift them. However, due to the recently proven no-go theorem, such a tiny uplifting cannot be achieved by a small modification of the parameters of the theory. We illustrate this general result by investigation of models with a single chiral superfield recently proposed by Ketov and Terada. We show that the addition of a small constant or a linear term to the superpotential of a model with a stable supersymmetric Minkowski vacuum converts it to an AdS vacuum, which results in a rapid cosmological collapse. One can avoid this problem and uplift a supersymmetric Minkowski vacuum to a dS vacuum with $V_{0}\sim 10^{-120}$ without violating the no-go theorem by making these extra terms large enough. However, we show that this leads to a strong supersymmetry breaking in the uplifted vacua.  

\end{abstract}

\vspace{24pt}
\end{titlepage}



\section{Introduction}

The embedding of inflationary scenarios in supergravity has been studied for over three decades by now. The first supergravity implementation of the chaotic inflation scenario in supergravity was proposed back in 1983 \cite{Goncharov:1983mw}, just a few months after the invention of the general chaotic inflation paradigm  \cite{Linde:1983gd}. It was based on the theory of a single chiral superfield with a plateau-type inflaton potential. Its predictions perfectly match the presently available observational data.

A subsequent development of inflation in supergravity shifted in the direction of flexibility of the models. The models offering a functional freedom of the choice of the inflaton potentials have been proposed only relatively recently, see e.g.  \cite{Kawasaki:2000yn,Kallosh:2010xz,Ferrara:2013rsa,Ferrara:2014kva,Kallosh:2014via}. 
Most  of the new models involved more than one chiral multiplet. For example, one may start with a broad class  of  theories    \cite{Kawasaki:2000yn,Kallosh:2010xz} of  two fields, $S= s\, e^{i\theta}/\sqrt 2$ and  $\Phi= (\phi+i\chi)/\sqrt 2$, with the superpotential
${W}= Sf(\Phi)$,
where $f(\Phi)$ is a real holomorphic function such that  $\bar f(\Phi) = f(\Phi)$.  The \K \, potential can be  chosen to have functional form
$K= K((\Phi-\bar\Phi)^2,S\bar S).$
In this case, the \K\, potential does not depend on $\phi$. The potential energy $V$ in this class of models has an extremum at $s = \chi = 0$. If this extremum is in fact a minimum, which can be achieved by a proper choice of the \K\ potential, inflation occurs along the direction $s = \chi = 0$, and the field $\phi$ plays the role of the canonically normalized inflaton field with the potential 
$V(\phi) = |f(\phi/\sqrt 2)|^{2}.$
 All scalar fields have canonical kinetic terms along the inflationary trajectory $S = {\rm Im}\, \Phi = 0$.  Alternatively, one can take $
K= K((\Phi+\bar\Phi)^2,S\bar S)$ and $W= S f(-i \Phi)$.
Then, the \K\, potential does not depend on $\chi$, which plays the role of the inflaton field with the potential $V(\chi) = |f(\chi/\sqrt 2)|^{2}$. 
One may also construct successful inflationary models using logarithmic potentials such as $K= -3\, \alpha \log (T + \bar T - C \bar C)$  \cite{Cecotti:1987sa,Kallosh:2013lkr,Ellis:2013nxa,Roest:2013aoa,Kallosh:2013yoa,Cecotti:2014ipa}.

In the cases described above, we have 2 superfields, with 4 scalar degrees of freedom, $s$, $\theta$, $\phi$ and $\chi$, but only one of them, corresponding to the inflaton field, participates in the cosmological evolution. Due to the consistency of this truncation, it is simple to investigate inflation in this scenario, and one can easily find functions $f(\Phi)$ which yield any desirable values of the cosmological parameters $n_{s}$ and $r$. In some models, one should take special care of stability of the field $S$ at $s =0$. It can be done either by adding terms $(S\bar S)^{2}$ to the \K\ potential  \cite{Kallosh:2010xz}, or by using a nilpotent chiral field $S$ which does not have scalar degrees of freedom  \cite{Ferrara:2014kva}. In this last case, the cosmological evolution is described by a single unconstrained chiral superfield $\Phi$. Moreover, one can generalize these inflationary models in such a way that they will also describe the present stage of the cosmological acceleration with a tiny cosmological constant $V_{0} \sim 10^{{-120}}$ and with a flexible scale of supersymmetry breaking, without introducing additional unconstrained chiral superfields \cite{Kallosh:2014via,Dall'Agata:2014oka,Kallosh:2014hxa,Linde:2014hfa}.

Recently, a new class of inflationary theories with a single chiral superfield $\Phi$ was proposed by Ketov and Terada (KT) \cite{Ketov:2014qha,Ketov:2014hya}. Following \cite{Ketov:2014hya}, one may consider a logarithmic \K\, potential of the form
\be\label{KK}
 K= -3\ln \left[ 1 + \frac{\Phi + \bar{\Phi} + \zeta \left( \Phi + \bar{\Phi}\right)^4}{\sqrt{3}}\right]\,.
\ee
The term with constant parameter $\zeta$ serves to stabilize the field $\chi$ during inflation at  $\chi \approx 0$. The main idea is that by making $\zeta$ sufficiently large one can make the field component $\phi$ heavy and constrained to a very small range of its values, 
$\phi \ll 1$, so it plays almost no role during inflation with the inflaton field $\chi \gg 1$. 
For superpotentials
\be
W = {1\over \sqrt 2} f(-\sqrt 2 i\Phi) \,,
\ee
where $f$ is a real function of its argument, the potential along the inflaton direction $\phi \ll 1$ becomes
\be\label{quadr}
V \approx \bigl[f'(\chi)\bigr]^{2} \ .
\ee
For example, for $W = {1\over 2}\Phi^{2}$ one recovers the simplest chaotic inflation potential $V = {m^{2}\over 2}\chi^{2}$ along the direction $\phi = 0$. A numerical investigation of this scenario in \cite{Ketov:2014hya} confirms that for sufficiently large $\zeta$, the field $\phi$ practically vanishes during the main part of inflation. Its evolution begins only at the very end of inflation, so the cosmological predictions almost exactly coincide with the predictions of the quadratic scenario. At the end of inflation, the field rolls down towards its supersymmetric Minkowski vacuum at $\Phi =0$, where $V =0$, $W =0$, and supersymmetry is restored. 

A similar conclusion is indeed valid for a large variety of superpotentials $W(\Phi)$, but not for all of them. In particular, we will show that one can have a consistent inflationary scenario in the theory with the simplest superpotential $W = c\Phi + d$, but both fields $\phi$ and $\chi$ evolve and play an important role. At the end of inflation, the field may roll to a Minkowski vacuum with $V = 0$ or to a dS vacuum with a tiny cosmological constant $V \sim 10^{{-120}}$. This is an encouraging result, since a complete cosmological model must include both the stage of inflation and the present stage of acceleration of the universe, and our simple model with a linear potential successfully achieves it. However, this success comes at a price: in this model, supersymmetry after inflation is strongly broken and the gravitino mass is  $2\times  10^{13}$ GeV, which is much greater than the often assumed TeV mass range. 

In view of this result, one may wonder what will happen if one adds a tiny correction term $c\Phi +d$ to the superpotential of the inflationary models described in \cite{Ketov:2014hya} with supersymmetric Minkowski vacua. Naively, one could expect that, by a proper choice of small complex numbers $c$ and $d$, one can easily interpolate between the AdS, Minkowski and dS minima. In particular, one could think that for small enough values of these parameters, one can conveniently fine-tune the value of the vacuum energy, uplifting the original supersymmetric minimum to the desirable dS vacuum energy with $V \sim 10^{{-120}}$.

However, the actual situation is very different. We will show that adding a small term $c\Phi +d$ always shifts the original Minkowski minimum down to AdS, which does not correctly describe our world. Moreover, unless the parameters $c$ and $d$ are exponentially small, the negative cosmological constant in the AdS minimum leads to a rapid collapse of the universe. For example, adding a tiny constant $d \sim 10^{{-54}}$ leads to a collapse within a time scale much shorter than its present age.  Thus, the cosmological predictions of the models of \cite{Ketov:2014hya} with one chiral superfield and a supersymmetric Minkowski vacuum are incredibly unstable with respect to even very tiny changes of the superpotential. Of course one could forbid such terms as $c\Phi +d$ by some symmetry requirements, but this would not address the necessity to uplift the Minkowski vacuum to $V \sim 10^{{-120}}$.

While we will illustrate this surprising result using KT models as an example, the final conclusion is very general and valid for a much broader class of theories with a supersymmetric Minkowski vacuum; see a discussion of a related issue in \cite{Kallosh:2014via}. We will show that this result is a consequence of the recently established no-go theorem of \cite{Kallosh:2014oja} (see also \cite{GomezReino:2006dk, Hardeman:2010fh}), which is valid for arbitrary \K\ potentials and superpotentials and also applies in the presence of multiple superfields:
\begin{quote}
\it One cannot deform a stable supersymmetric Minkowski vacuum with a positive definite mass matrix to a non-supersymmetric de Sitter vacuum by an infinitesimal change of the \K\, potential  and superpotential.
\end{quote}
This no-go theorem can be understood from the role of the sGoldstino field, the scalar superpartner of the would-be Goldstino spin-1/2 field (as also emphasized in \cite{Covi:2008ea,Covi:2008cn,Achucarro:2012hg}). Since the mass of the sGoldstino is  set by the order parameter of supersymmetry breaking, it must vanish in the limit where supersymmetry is restored. The only SUSY Minkowski vacua that are continuously connected to a branch of non-supersymmetric extrema therefore necessarily have a flat direction to start with: this is the scalar field that will play the role of the sGoldstino after spontaneous SUSY breaking. A corollary of this theorem is that one cannot obtain a dS vacuum from a stable SUSY Minkowski vacuum by a small deformation. As we will see, this is exactly what forbids a small positive CC after an infinitesimal change of the KT starting point.

As often happens, the no-go theorem does not mean that uplifting of the supersymmetric Minkowski minimum to a dS minimum is impossible. In order to achieve that, the modification of the superpotential should be substantial. We will show how one can do it, thus giving a detailed illustration of how this no-go theorem works and how one can overcome its conclusions by changing the parameters of the correction term $c\Phi +d$ beyond certain critical values. For example, one can take $d =0$ and slowly increase $c$. For small values of $c$, the absolute minimum of the potential corresponds to a supersymmetric AdS vacuum. When the parameter $c$ reaches a certain critical value, the minimum of the potential ceases to be supersymmetric, but it is still AdS. With a further increase of $c$, the minimum is uplifted and becomes a non-supersymmetric dS vacuum state. Once again, we will find that the modification of the superpotential required for the tiny uplifting of the vacuum energy  by $V_{0} \sim 10^{{-120}}$ leads to a strong supersymmetry breaking, with the gravitino mass many orders of magnitude greater than what is usually expected in supergravity phenomenology.

This problem can be solved by introducing additional chiral superfields responsible for uplifting and supersymmetry breaking. However,  this may require an investigation of inflationary evolution of multiple scalar fields, unless the additional fields are strongly stabilized \cite{Dudas:2012wi} or belong to nilpotent chiral multiplets \cite{Ferrara:2014kva,Kallosh:2014via,Dall'Agata:2014oka,Kallosh:2014hxa,Linde:2014hfa}. 

The outline of this paper is as follows. In section 2 we will discuss the possibility to realize inflation and dark energy with a \K\ potential \eqref{KK} supplemented with a superpotential consisting of a constant and linear part. This is generalized to include a quadratic term in section 3. We conclude in section 4, while in the appendix we elucidate the vacuum structure of the model studied in section 2.

\section{Inflation and uplifting with a linear superpotential}\label{Wlin}

\begin{figure}[h!]
\begin{center}
\includegraphics[width=7cm]{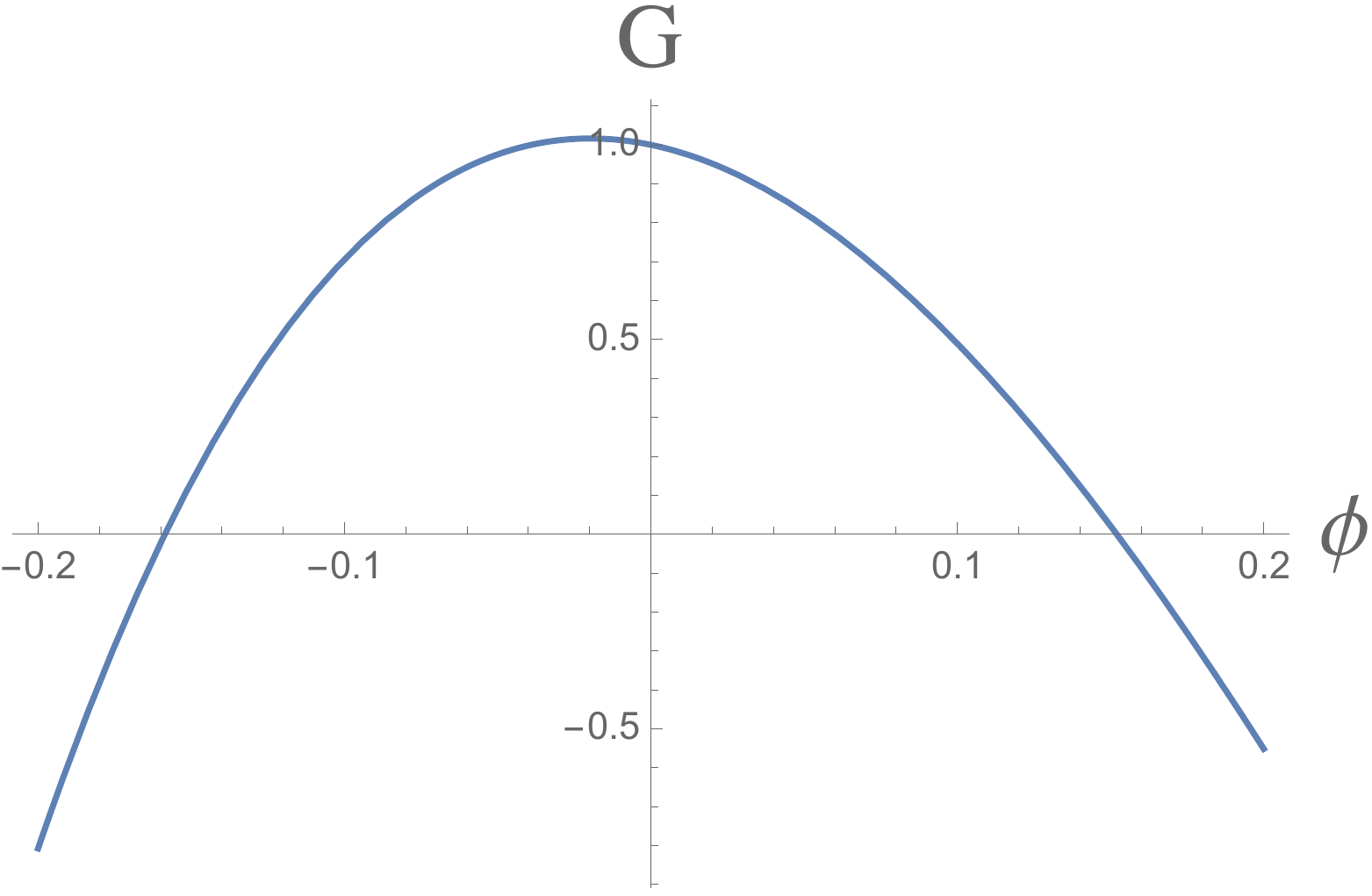}
\caption{\footnotesize The coefficient in front of the kinetic term for the field $\Phi$ as a function of $\phi$ for  $\zeta =1$.}\label{1}
\end{center}
\vspace{-0.5cm}
\end{figure}

To understand the basic features of the theories with the \K\ potential \eqref{KK}, it is instructive to calculate the coefficient $G(\phi,\chi)$ in front of the kinetic term of the field $\Phi$. For an arbitrary choice of the superpotential, this coefficient is given by
\be
G(\phi,\chi)= {3 (1 + 32 \zeta^2 \phi^6 - 8 \zeta \phi^2 (3 \sqrt{3} + \sqrt{2} \phi))\over(\sqrt 3 + 
     \sqrt{2} \phi + 4  \zeta \phi^4)^2} \ .
\ee
This function does not depend on $\chi$. For small $\phi$ the fields are canonically normalized. $G(\phi,\chi)$ is positive at small $\phi$, while it vanishes and becomes negative for larger values of $|\phi|$ (provided $\zeta > 0$). Thus the kinetic term is positive definite only in a certain range of its values, depending on the constant $\zeta$. In this paper, we will usually take $\zeta =1$, to simplify the comparison with  \cite{Ketov:2014hya}, see Fig. \ref{1}.
It is equally important that the expression for the potential $V$ in this theory, for any superpotential, contains the  coefficient $1 + 32 \zeta^2 \phi^6 - 8 \zeta \phi^2 (3 \sqrt{3} + \sqrt{2} \phi)$ in the denominator, so it becomes infinitely large exactly at the boundaries of the moduli space where the kinetic term vanishes (for $\zeta=1$, the boundaries are located at $\phi\approx \pm 0.15$). For large $\zeta$, the domain where $G$ is positive definite becomes more and more narrow, which is why the field $\phi$ becomes confined in a narrow interval, whereas the field $\chi$ is free to move and play the role of the inflaton field. This is very similar to the mechanism of realization of chaotic inflation proposed earlier in a different context in Section 4 of \cite{Kallosh:2014qta}.

\begin{figure}[htb]
\begin{center}
\includegraphics[width=10cm]{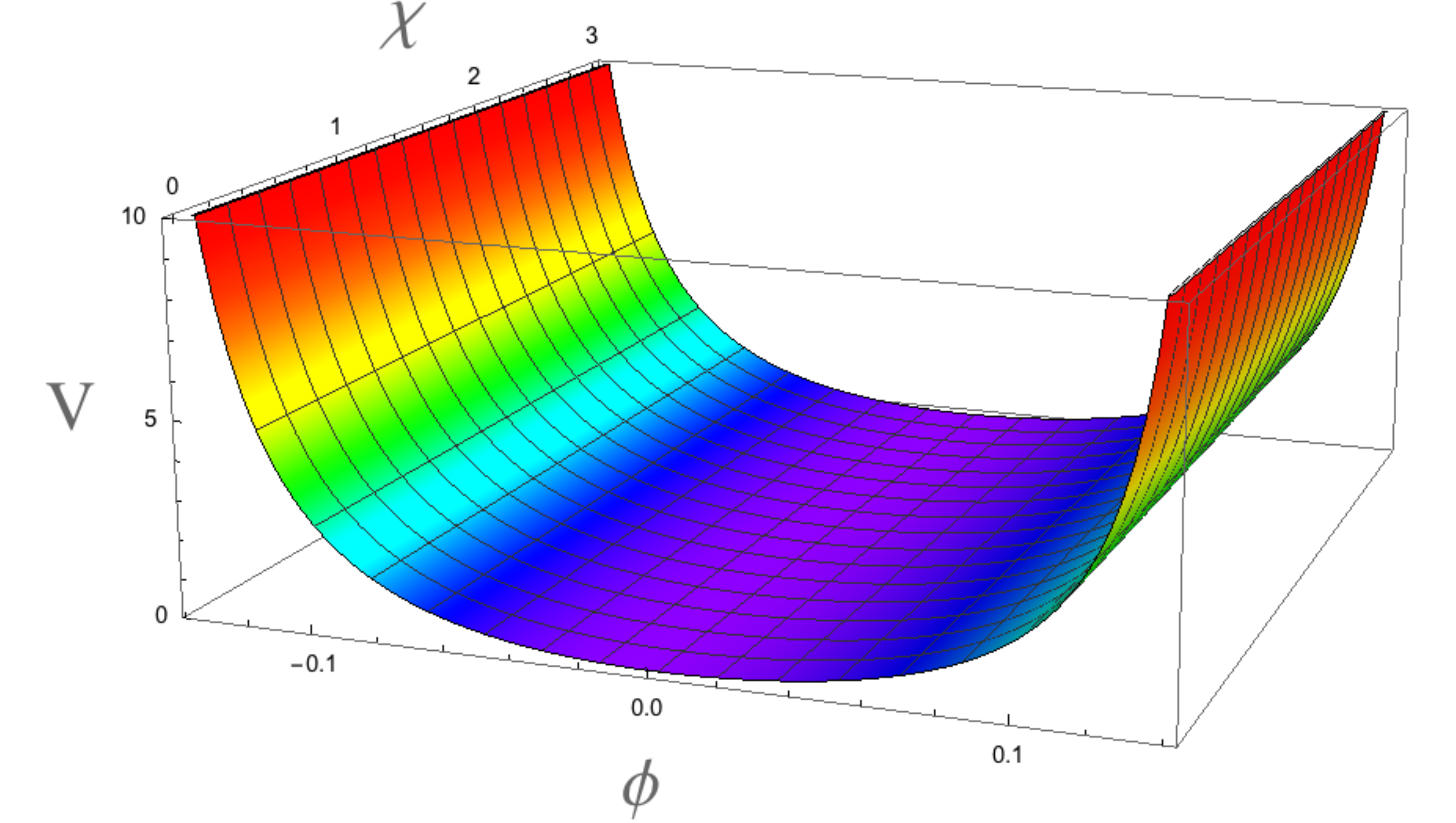}
\caption{\footnotesize The scalar potential in the theory with a constant superpotential $W = m$. For $\zeta = 1$, it blows up at $\phi \approx 0.15$, and it does not depend on the field $\chi$, forming a narrow Minkowski valley surrounded by infinitely steep walls.}\label{2}
\end{center}
\vspace{-0.5cm}
\end{figure}

We will study inflation in this class of theories by giving some examples, starting from the simplest ones. The simplest superpotential to consider is a  constant one, $W = m$. In this case, the potential does not depend on the field $\chi$. It blows up, as it should, at sufficiently large $\phi$, and it vanishes at $\phi = 0$, see Fig. \ref{2}. This potential does not describe inflation.

\begin{figure}[htb]
\begin{center}
\includegraphics[width=10cm]{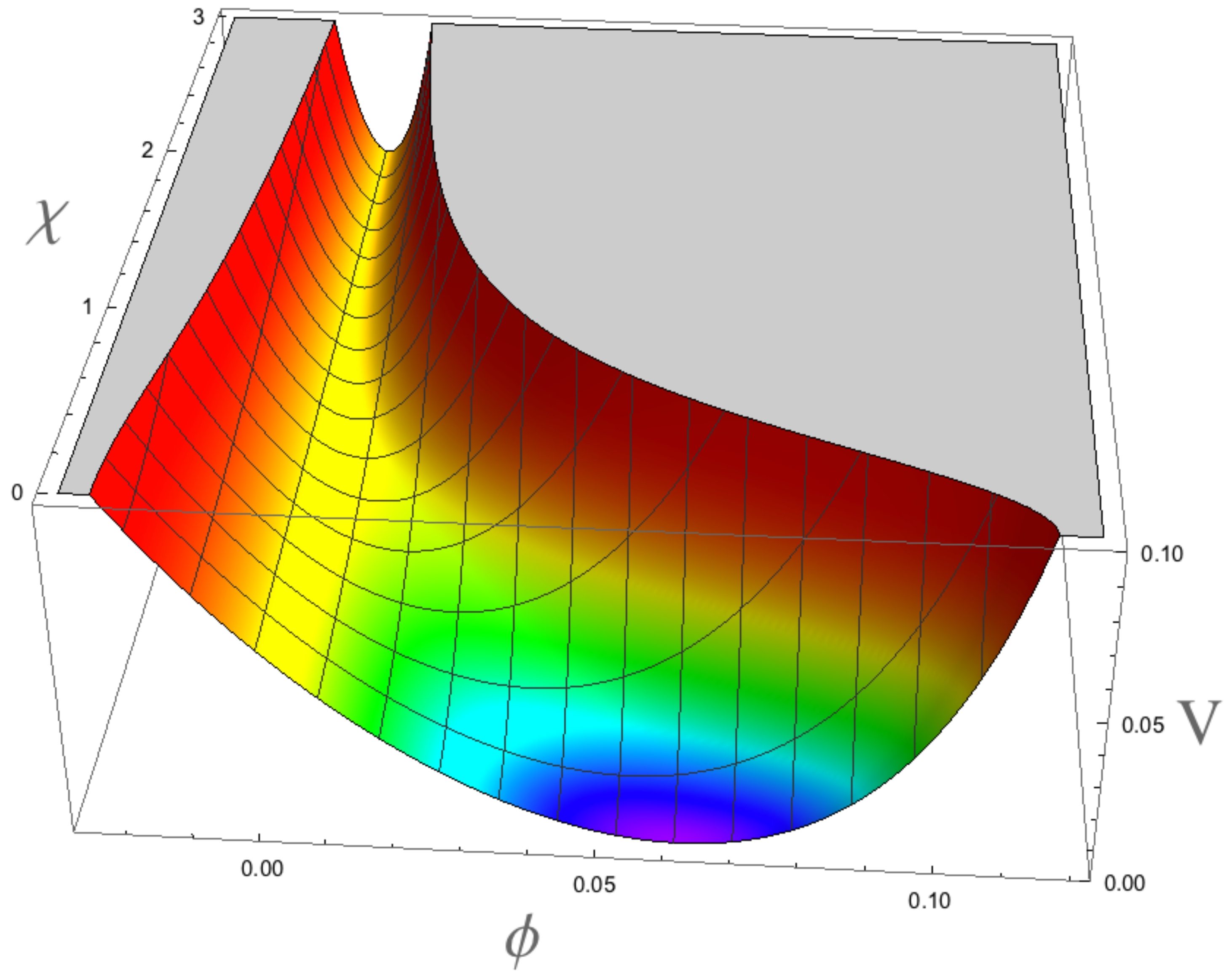}
\caption{\footnotesize The scalar potential in the theory with  $W = m\, (c\Phi +1)$, for $\zeta = 1$. For $c \approx 3.671$, it has a dS valley, and a near-Minkowski minimum at $\chi =0$, $\phi \approx 0.06$. Inflation happens when the field slowly moves along the nearly flat valley and then rolls down towards the minimum of the potential. It is a two-field inflation, which cannot be properly studied by assuming that $\phi = 0$ during the process.}\label{3}
\end{center}
\vspace{-0.5cm}
\end{figure}

As a next step, we will consider a superpotential with a linear term
\be
W = m\, (c\Phi +1) \ .
\ee
In this case, just as in the case considered above, the potential has an exactly flat direction at $\phi = 0$, but now the potential at  $\phi = 0$ is equal to
\be
V(\phi = 0,\chi) = m^{2}c\, (c-2\sqrt 3) \ .
\ee
Thus for $c < 2\sqrt 3$ it is an AdS valley, but for $c > 2\sqrt 3$ it is a dS valley. But this does not tell us the whole story. At large $\chi$, the minimum of the potential in the $\phi$ direction is approximately at $\phi = 0$, but at smaller $\chi$, the minimum shifts towards positive $\phi$. For\footnote{An understanding of this value of $c$ and its role in terms of (non-)supersymmetric branches is given in appendix A.} $c \approx 3.671$, the potential has a global non-SUSY Minkowski minimum with $V =0$ at $\chi =0$ and $\phi \approx 0.06$. By a minuscule change of $c$ one can easily adjust the potential to have the desirable value $V_{0} \sim 10^{{-120}}$ at the minimum. This requires fine-tuning, but it should not be a major problem in the string landscape scenario. The full potential is shown in Fig.~\ref{3}.

\begin{figure}[htb]
\begin{center}
\includegraphics[width=8cm]{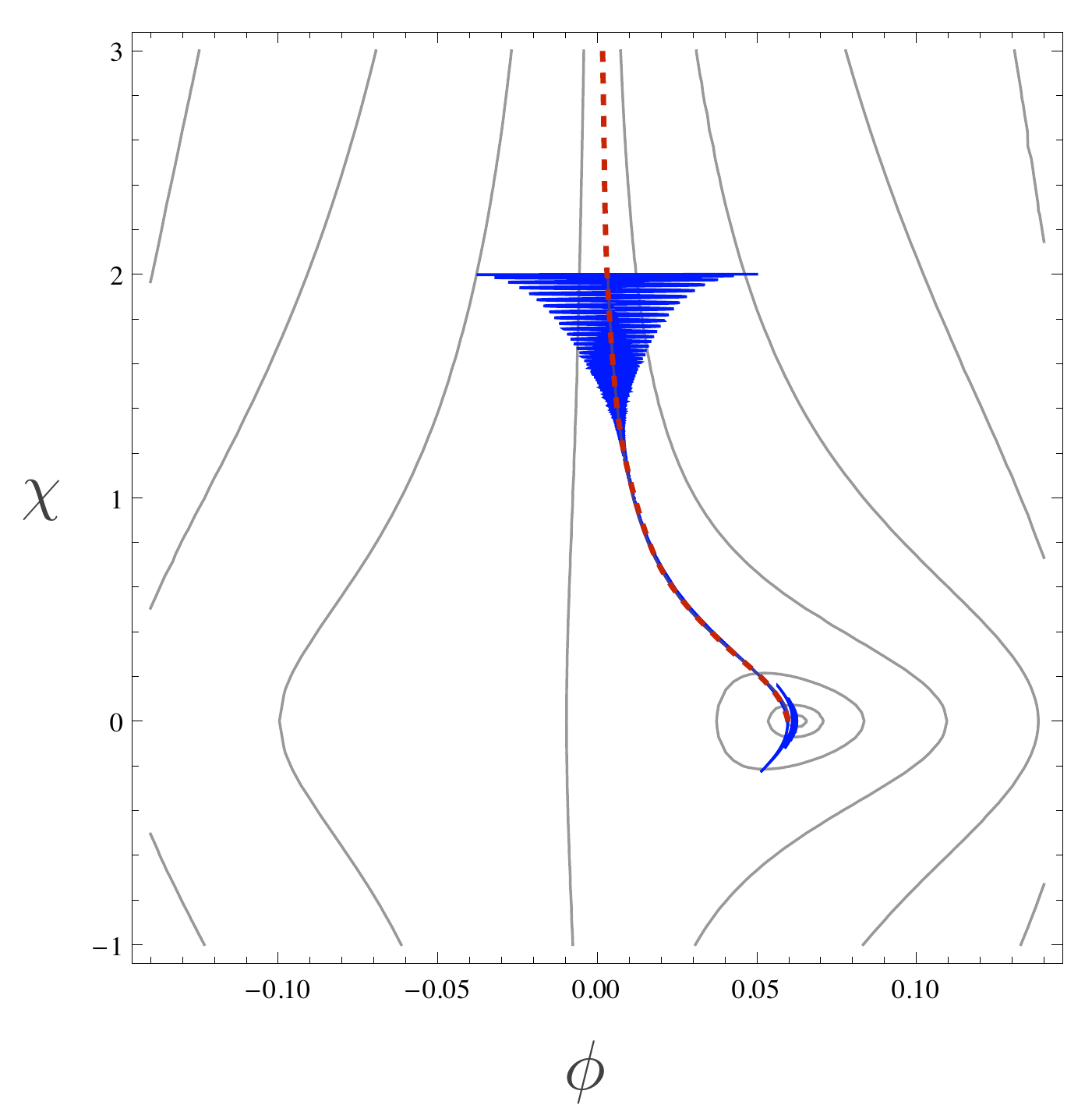}
\caption{\footnotesize The dynamical evolution of the inflaton field (blue line) in the model with $W=m(c\Phi+1)$, for $\zeta=1$. The adiabatic approximation of the effective potential (dashed red line) and the contour plot of $V(\phi,\chi)$ in logarithmic scale are shown as superimposed. There is an initial stage of oscillations before the field approaches the inflationary attractor, as well as the final stage of post-inflationary oscillations. However, during inflation, which happens between these two oscillatory stages, the field accurately follows the position of the adiabatically changing minimum of the potential $V(\phi(\chi),\chi)$. }\label{Ev}
\end{center}
\vspace{-0.5cm}
\end{figure}

Inflation in this models happens when the field slowly moves along the nearly flat valley and then rolls down towards the minimum of the potential. It is a two-field dynamics, which cannot be properly studied by assuming that $\phi = 0$ during the process, as proposed in  \cite{Ketov:2014qha,Ketov:2014hya}. Indeed, the potential along the direction $\phi = 0$ is exactly constant, so the field would not even move if we assumed that during its motion. However, because of the large curvature of the potential in the $\phi$ direction, during inflation this field rapidly reaches an inflationary  attractor trajectory and then adiabatically follows the position of the  minimum of the  potential $V(\phi,\chi)$ for any given value of the field $\chi(t)$. This can be confirmed by a numerical investigation of the combined evolution of the two fields  whose dynamics is shown in Fig.~\ref{Ev}. 

Then, the adiabatic approximation of the effective potential driving inflation reads 
\be
V(\phi(\chi), \chi) = m^{2}c\, (c-2\sqrt 3) -\frac{2m^2(c-\sqrt{3})^2}{27\sqrt{3}\chi^2}\,,
\ee
neglecting higher order terms which play no role in the inflationary plateau. The effective fall-off of $1/\chi^2$ is responsible for determining the main properties of a fully acceptable inflationary scenario.

This investigation shows that this simplest model leads to a desirable amplitude of inflationary perturbations for $m \sim 7.75 \times 10^{-6}$, in Planck units. The inflationary parameters $n_{s}$ and $r$ in this model are given by (at leading order in $1/N$)
 \begin{align}
   n_{s} = 1 - \frac{3}{2N} \,, \quad r = { 2(c-\sqrt3)\over \sqrt{26 c (\sqrt{3} c -6)} \, N^{3/2}} \,.
  \end{align}
Numerically, we find $n_s \approx 0.975$ and $r \approx 0.0014$ for $N=60$, in excellent agreement with the leading $1/N$ approximation. We checked that the values of $n_{s}$ remains approximately the same in a broad range of $\zeta$, from $\zeta = 0.1$ to $\zeta = 10$. The value of the parameter $r$ slightly changes but remains in the $10^{{-3}}$ range. As of now, all of these outcomes are in good agreement with the Planck data.

However, this simplest inflationary model has a property which is shared by all other models of this class to be discussed in this paper: supersymmetry is strongly broken in the minimum of the potential. In particular, for $\zeta = 1$, the superpotential at the minimum is given by $W\approx 9 \times 10^{-6}$, and the gravitino mass is $m_{3/2} \sim 8.34 \times 10^{-6}$, in Planck units, i.e.   $m_{3/2} \sim 2 \times 10^{13}$ GeV. This is many orders of magnitude higher than the gravitino mass postulated in many phenomenological models based on supergravity.

Of course, supersymmetry may indeed be broken at a very high scale, but nevertheless this observation is somewhat worrisome. One could expect that this is a consequence of the simplicity of the model that we decided to study, but we will see that this result is quite generic.

\section{Inflation and uplifting with a quadratic superpotential}\label{secquad}

As a second example, we will discuss the next simplest model, defined by 
\be\label{Wquad}
W=\tfrac{1}{2} m \Phi^2\,.
\ee
This case was one of the focuses of \cite{Ketov:2014hya} and gives rise to a quadratic inflationary potential. As we will demonstrate, perturbing such a superpotential by means of a linear and constant term, leads to general properties which are shared by the class discussed in the previous section.

\begin{figure}[htb]
\begin{center}
\includegraphics[width=8cm]{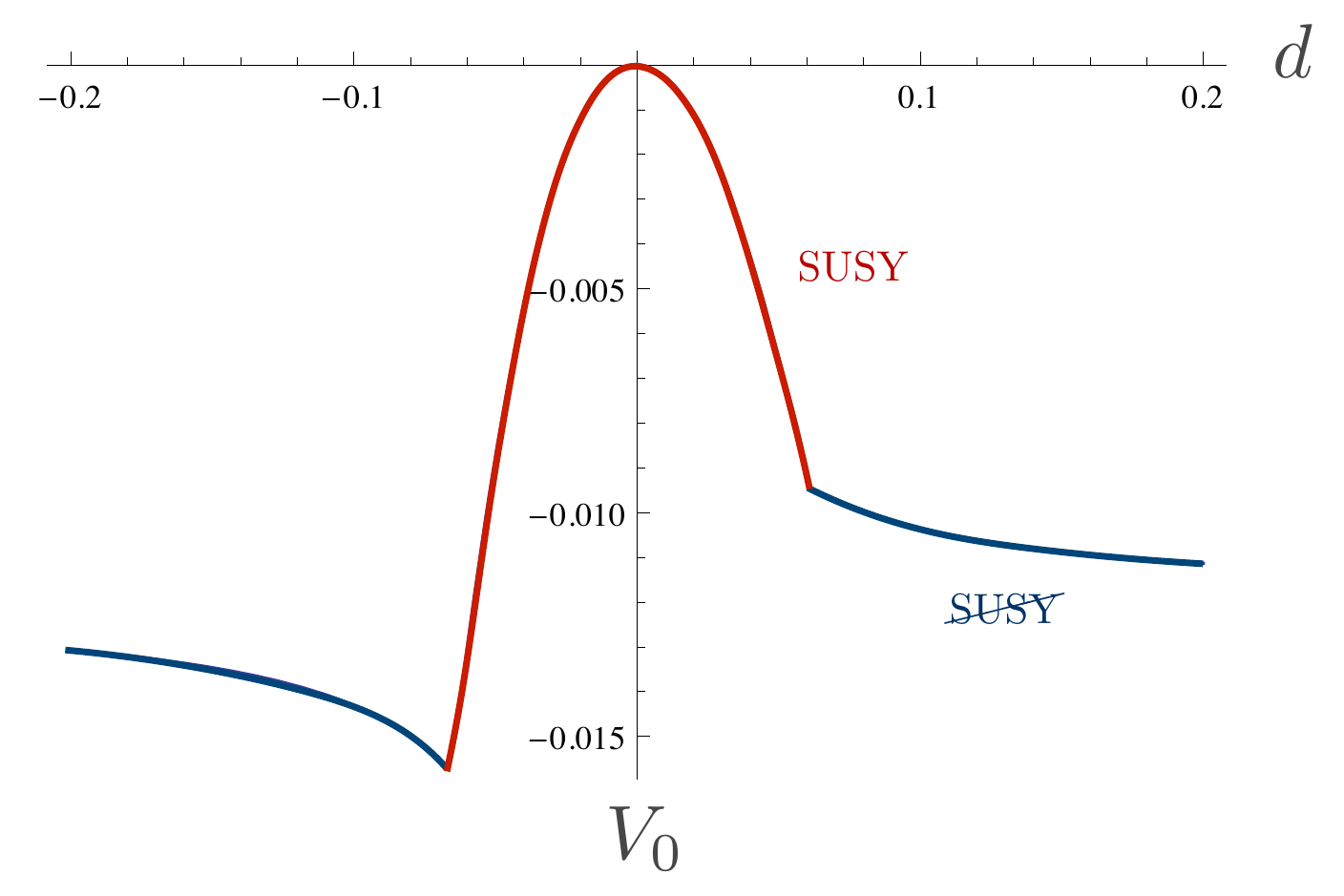}
\includegraphics[width=8cm]{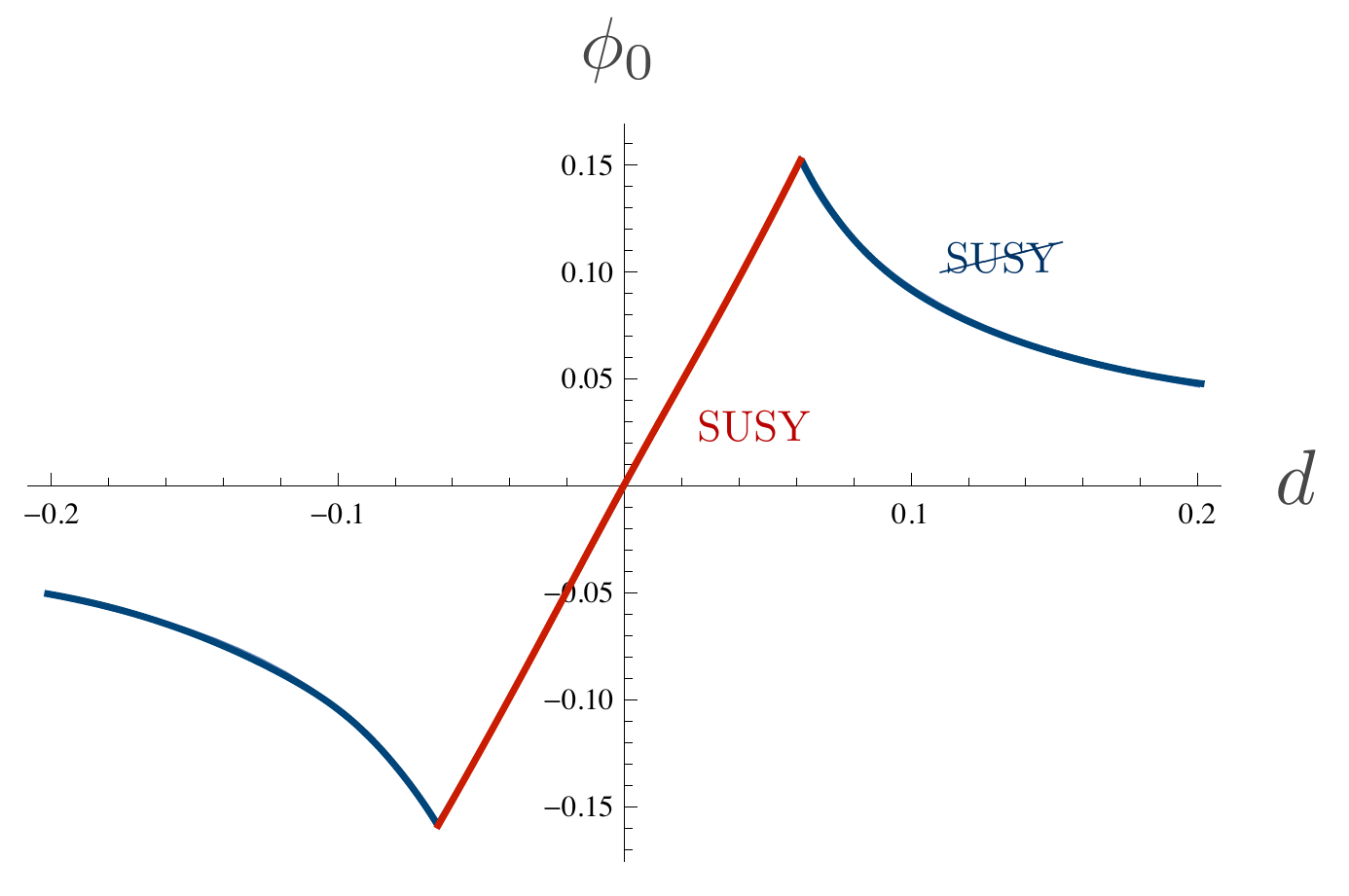}
\caption{\footnotesize The value of the cosmological constant (left panel) in the minimum and its location $\phi_0$ (right panel) as a function of the constant term $d$ in the superpotential \eqref{Wd}. The two branches of solutions (SUSY and non-SUSY), within the fundamental physical domain $|\phi|\lesssim0.15$, are shown in different colors.  At larger (positive or negative) values of the constant, both the CC and the location $\phi_0$ level off to a constant. Plots obtained for $m = \zeta = 1$.}\label{small-const}
\end{center}
\vspace{-0.5cm}
\end{figure}

We will start by perturbing this model via a constant term such as
\be\label{Wd}
W= m \left(\tfrac{1}{2}\Phi^2 +d\right)\,.
\ee
The inflationary regime is unaffected by such correction and the scalar potential still reads $V=\tfrac{1}{2}m^2\chi^2$, at $\phi=0$. However, the vacuum of $V(\phi,\chi)$ will move away from the supersymmetric Minkowski minimum, originally placed at $\Phi=0$, but just in the $\phi$-direction (because the superpotential is symmetric). Then, for small parameter values, the minimum of $\phi$ moves as
\be
\phi_0= \sqrt{6}d-\sqrt{\frac{3}{2}}d^2\,.
\ee
This shift immediately leads to an AdS phase which, at small values of $d$, goes as
\be
V_0= -\sqrt{3}m^2d^2\,,
\ee
which is fully in line with the no-go theorem \cite{Kallosh:2014oja} summarized in the Introduction. These solutions do not break supersymmetry and they can be obtained by the equation $\mathcal{D}_{\Phi} W=0$. As $|d|$ increases, such a SUSY vacuum moves further away from the origin and, at one point, it crosses the singular boundary of the moduli space. Then, if we search for numerical solutions within the strip corresponding to the correct sign of the kinetic terms (this means for $|\phi|\lesssim0.15$), we run into a feature which will be common also in other examples: for specific values of $d$, the SUSY-branch of vacuum solutions leaves the fundamental physical domain $|\phi|\lesssim0.15$ and it is replaced by a new branch of vacua with broken supersymmetry. This is shown in Fig.~\ref{small-const}. However, as one keeps increasing the absolute value of $d$, $\phi_0$ approaches a constant value which corresponds to an asymptotic AdS phase. Therefore, perturbing $W$ by means of a constant term does not help to uplift to dS.

As second step, we include a linear correction such that the superpotential reads
\be\label{Wc}
W= m \left(\tfrac{1}{2}\Phi^2 +c\Phi\right)\,.
\ee
where the coefficients are real due to the constraint on\footnote{Perturbing the superpotential by means of a linear term with imaginary coefficient such as $ic\Phi$ is equivalent to adding a positive constant $c^2$. This is a direct consequence of the shift symmetry of the \K\ potential.} $W$.

\begin{figure}[htb]
\begin{center}
\includegraphics[width=8cm]{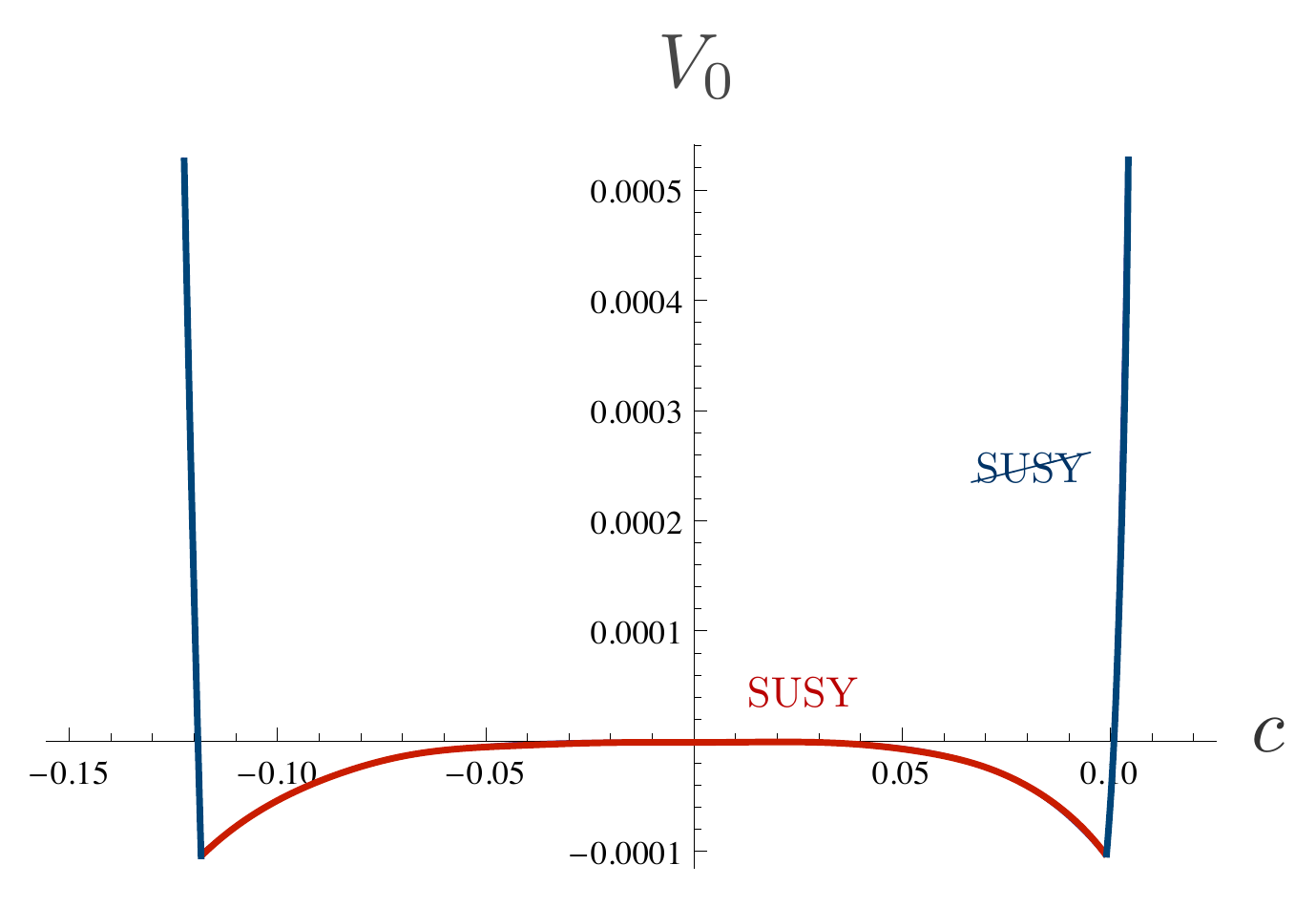}
\includegraphics[width=8cm]{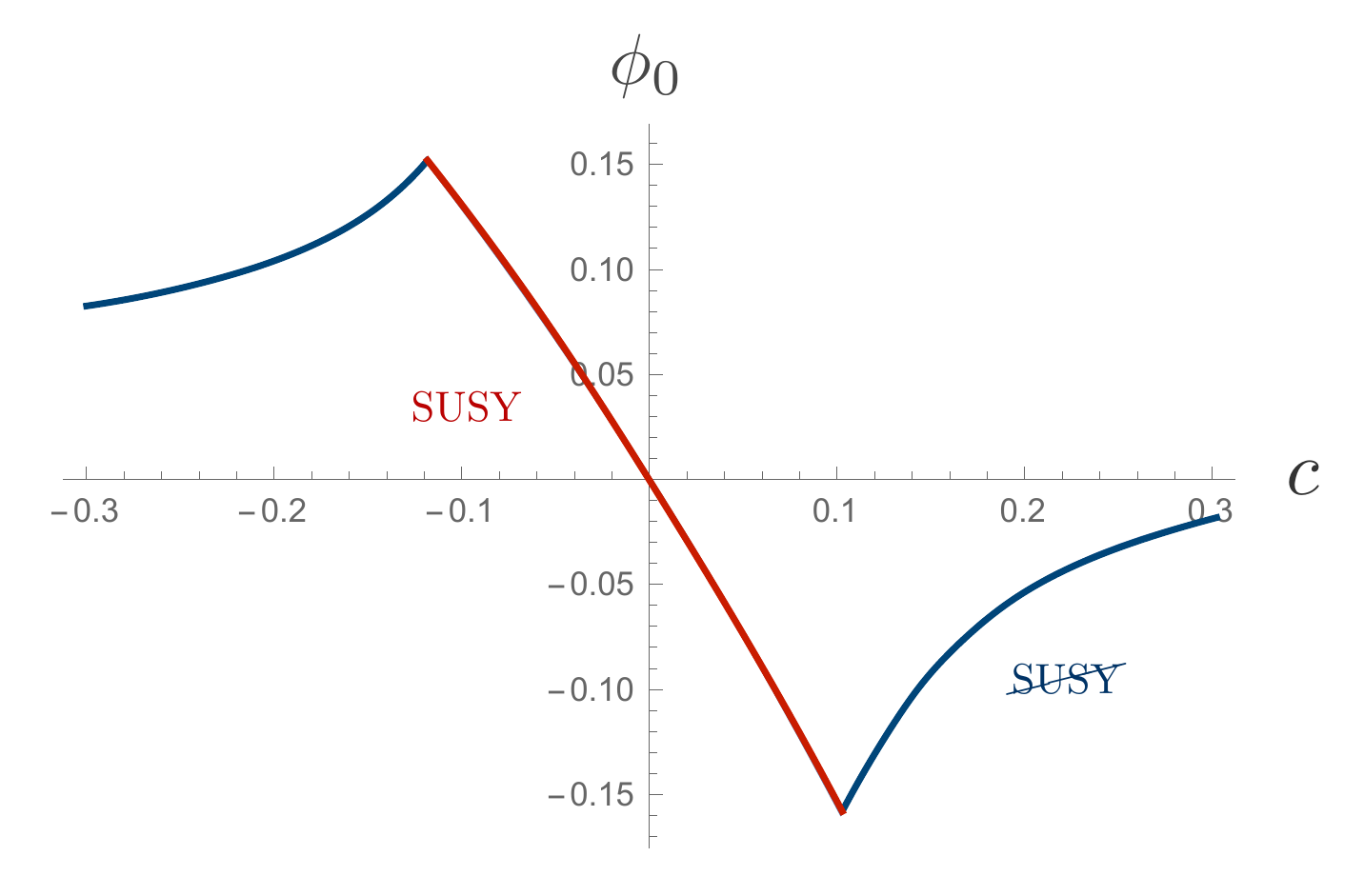}
\caption{\footnotesize The value of the cosmological constant (left panel) in the minimum and its location $\phi_0$ (right panel) as a function of the linear term in the superpotential. The two branches of solutions (SUSY and non-SUSY), within the fundamental physical domain $|\phi|\lesssim0.15$, are shown in different colors. At larger (positive or negative) values of the coefficient $c$, the location $\phi_0$ levels off to a constant while the CC approaches a quadratic shape. Plots obtained for $m = \zeta = 1$.}\label{small-lin}
\end{center}
\vspace{-0.5cm}
\end{figure}

Similarly to the previous case, the SUSY Minkowski vacuum is perturbed by such correction and, at lowest order in $c$, it moves in the $\phi$-direction as
\be
\phi_0=-\sqrt{2}c-\sqrt{\frac{3}{2}}c^2\,,
\ee
leading to a vacuum energy given by
\be
V_0= -\sqrt{\frac{3}{4}}m^2c^4\,,
\ee
Then also in this case, as $|c|$ increases, such supersymmetric solutions move towards the boundary $\phi\approx\pm0.15$ and cross it. At the same point in parameter space, a new branch of non-supersymmetric solutions appears and, remarkably, this results into a sharp increase of the scalar potential at the minimum. In fact, this very quickly gives rise to a transition from AdS to dS, as it is shown in Fig.~\ref{small-lin}. 

The exact values for which these transitions happen are as follows. The transition from SUSY to non-SUSY vacua occurs at (calculated for $m= \zeta = 1$)
 \begin{align}
  c =  -0.118162 \,, \qquad c = 0.101918  \,,
 \end{align}
while the CC crosses through Minkowski at
 \begin{align}\label{MinkLin}
  c =  -0.119318 \,, \qquad c = 0.102692 \,.
 \end{align}

Note that, at finite $c$ values, the scalar potential passes through Minkowski. In contrast to the ground state at $c=0$, the new Minkowski vacua are non-supersymmetric, and hence can be deformed into dS without violating the no-go theorem. In fact, these non-supersymmetric Minkowski vacua are exactly the type of  structures that were identified in \cite{Kallosh:2014oja} as promising starting points for uplifts to De Sitter (although there the focus was on a hierarchy of supersymmetry breaking order parameters for different superfields). A minuscule deviation of $c$ from \eqref{MinkLin} will be sufficient to obtain the physical value of cosmological constant $V_0\sim10^{-120}$.

It is worthwhile to remark that the order of magnitude of the parameter $c$, for which we get a tiny uplifting to dS, is small with respect to the coefficient of the quadratic term in the superpotential \eqref{Wc}. This translates into the fact that the inflationary predictions will be basically unchanged with respect the simple scenario with a quadratic potential. In fact, the scalar potential in the direction $\phi=0$ reads
\be
V(\phi=0 ,\chi)= \frac{1}{2}\left(1-\sqrt{3}c\right)m^2\chi^2\ + m^2c^2\,.
\ee
At $\chi \lesssim O(1)$, the field $\phi$ no longer vanishes and starts moving towards the minimum of the potential. However, the main stage of inflation happens at $\chi \gg c =O(0.1)$, when $\phi$ nearly vanishes and the inflaton potential is approximately equal to $\frac{1}{2}\left(1-\sqrt{3}c\right)m^2\chi^2$. The main effect of this change of the potential is a slight change of normalization of the amplitude of the perturbations spectrum, which requires a small adjustment for the choice of the parameter $m$: 
\be
m\approx (6+5.2c)\cdot10^{-6}\,.
\ee

However, even though the inflationary regime is essentially unaffected by such a small correction, supersymmetry is strongly broken at the end of inflation, just as in the theory with a simple linear superpotential, discussed in Sec.~\ref{Wlin}. This is a direct consequence of the no-go theorem discussed above and of the impossibility of uplifting the SUSY Minkowski vacuum (corresponding to $c=0$) by an infinitesimal deformation of $W$. In particular, for values of $c$ leading to a realistic dS phase (these values are extremely close to \eqref{MinkLin}, corresponding to non-supersymmetric Minkowski) and  for $\zeta=1$, we obtain the following: for positive $c$, the superpotential at the minimum is $|W|\approx3.4\times 10^{-8}$ and the gravitino mass is $m_{3/2}\sim 4.2 \times 10^{-8}$, in Planck units, i.e. $m_{3/2}\sim 1.0 \times 10^{11}$ GeV; for negative $c$, the superpotential at the minimum is $|W|\approx3.8\times 10^{-8}$ and the gravitino mass is $m_{3/2}\sim 3.2 \times 10^{-8}$, in Planck units, i.e. $m_{3/2}\sim 7.6 \times 10^{10}$ GeV. These values are again well beyond the usual predictions of the low scale of supersymmetry breaking in supergravity phenomenology. 


\section{Discussion}\label{disc}


In this paper we have investigated the possibility to realize a model of inflation and dark energy in supergravity. As an example, we considered the class of single chiral superfield models proposed in \cite{Ketov:2014hya}. The models described in  \cite{Ketov:2014hya} share the following feature: The vacuum energy in these models vanishes, and supersymmetry is unbroken. One could expect that this is a wonderful first approximation to describe dS vacua with vanishingly small vacuum energy $V_{0}\sim 10^{{-120}}$ and small supersymmetry breaking with $m_{3/2} \sim 10^{-15}$ or $10^{-13}$ in Planck units. However, we have shown that this is not the case, because of the no-go theorem formulated in \cite{Kallosh:2014oja}. While it is possible to realize an inflationary scenario  that ends in a dS vacuum with $V_{0}\sim 10^{{-120}}$, these vacua cannot be infinitesimally uplifted by making small changes in the \K\ potential and superpotential. One can uplift a stable Minkowski with unbroken SUSY to a dS minimum, but it always requires large uplifting terms, resulting in a strong supersymmetry breaking with $m_{3/2}$ many orders of magnitude higher than the TeV  or even PeV range advocated by many supergravity phenomenologists. 

In our investigation, we also introduced a new model, which contained only linear and constant terms in the superpotential. This superpotential is simpler than those studied in \cite{Ketov:2014hya}, but we have found that this model does describe a consistent inflationary theory with  dS vacuum, which can have $V_{0}\sim 10^{{-120}}$. However, just as in all other cases considered in our paper, we found that supersymmetry is strongly broken after inflation in this model. While we have analyzed only some specific cases in detail, our conclusions apply to a much wider class of models, well beyond the specific models proposed in \cite{Ketov:2014hya}, because of the general nature of the no-go theorem of  \cite{Kallosh:2014oja}. 

Since there is no evidence of low scale supersymmetry at LHC as yet, one could argue that the large scale of supersymmetry breaking is not necessarily a real problem. However, it would be nice to have more flexibility in the model building, which would avoid this issue altogether. One way to get dS uplifting with small supersymmetry breaking, without violating the no-go theorem, is to add other chiral multiplets (e.g. Polonyi fields), and to strongly stabilize them to minimize their influence on the cosmological evolution, see e.g. \cite{Dudas:2012wi}. In certain cases, one can make the Polonyi fields so heavy and strongly stabilized that they do not change much during the cosmological evolution and do not lead to the infamous Polonyi field problem which bothered cosmologists for more than 30 years  \cite{Polonyi}.  A more radical approach, which allows to have a single scalar field evolution is to use the recently proposed models involving nilpotent chiral superfields \cite{Ferrara:2014kva,Kallosh:2014via,Dall'Agata:2014oka,Kallosh:2014hxa,Linde:2014hfa}, which have an interesting string theory interpretation in terms of D-branes \cite{Kallosh:2014wsa}. Adding the nilpotent superfields allows to uplift the vacuum energy and to achieve a controllable level of supersymmetry breaking.

\section*{Acknowledgments}

We are grateful to Renata Kallosh, Bert Vercnocke and  Ivonne Zavala for many useful discussions. AL is supported by the SITP and by the NSF Grant PHY-1316699 and by the Templeton foundation grant `Inflation, the Multiverse, and Holography.'  MS acknowledges financial support by the University of Groningen, within the Marco Polo Fund, and by COST Action MP1210. Finally, MS would like to thank the SITP for its warm hospitality.


\section*{Appendix: Vacuum structure of the linear superpotential}

In this appendix, we return to the simplest model with 
 \begin{align}
  W = m (c \Phi +1) \,,
 \end{align}
and explain how its vacuum structure can be understood in the same terms as that of the quadratic superpotential presented in Sec.~\ref{secquad}: the starting point is a SUSY vacuum, small deviations preserve SUSY and lower the CC, and only for large deviations one does break SUSY and the potential at the minimum can get zero or positive values. However, in contrast to the quadratic case, the SUSY range does not include $c=0$. 

\begin{figure}[htb]
\begin{center}
\includegraphics[width=8cm]{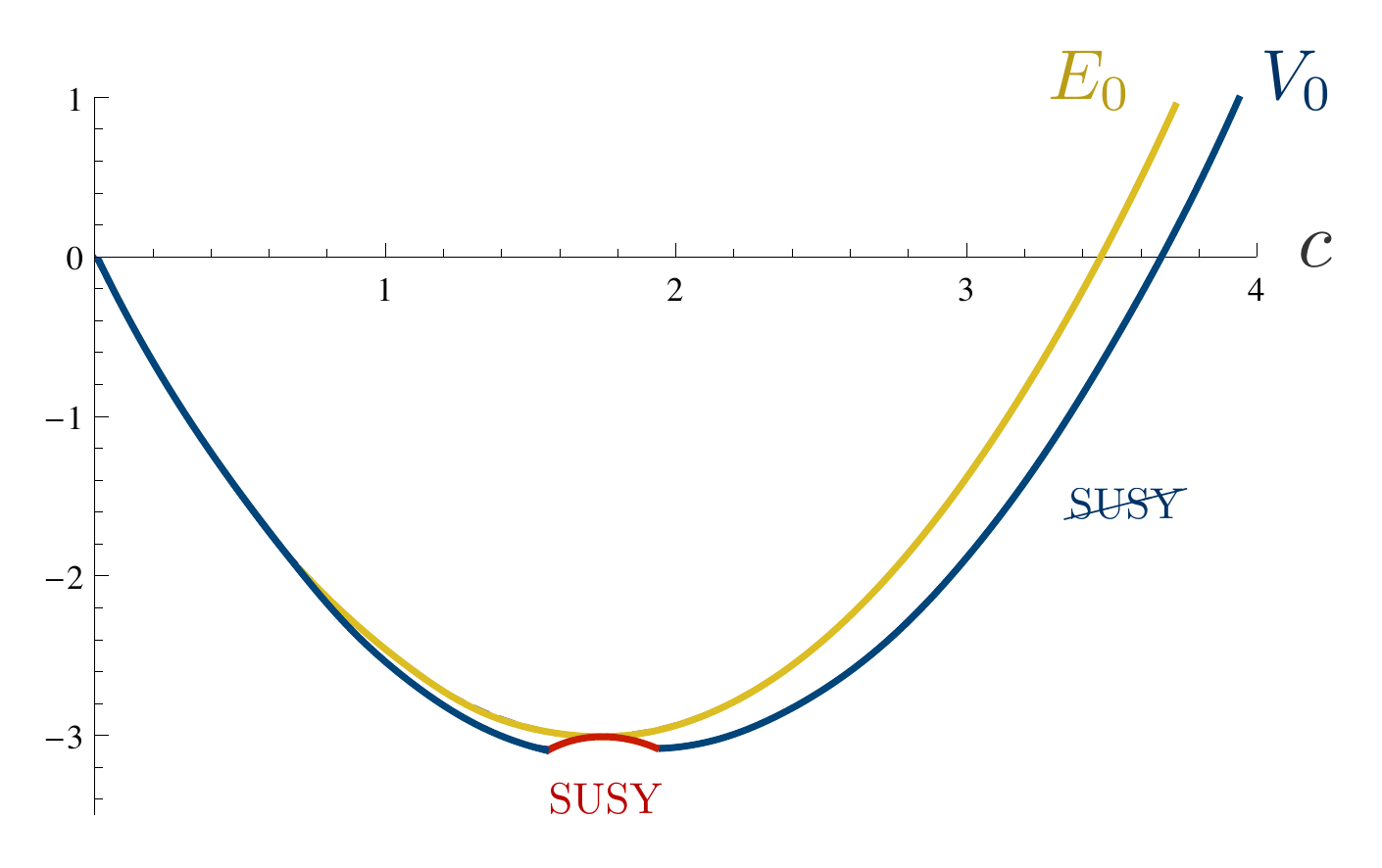}
\includegraphics[width=8cm]{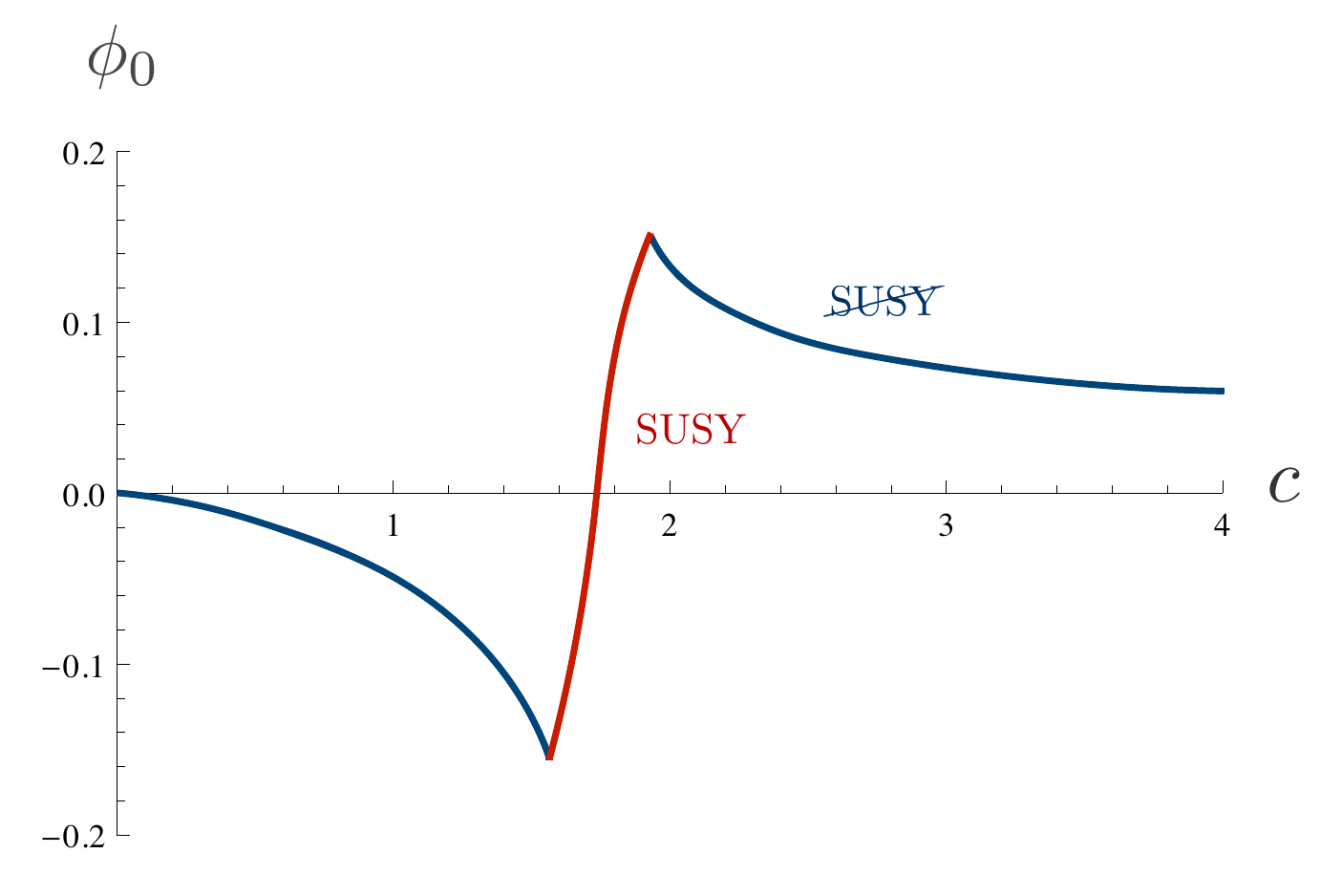}
\caption{\footnotesize The value of the cosmological constant (left panel) in the minimum and its location $\phi_0$ (right panel) as a function of the linear term in the superpotential.  The two branches of solutions (SUSY and non-SUSY), within the fundamental physical domain $|\phi|\lesssim0.15$, are shown in different colors.  The light yellow line in the left panel denotes the inflationary energy. Plots obtained for $m = \zeta = 1$.} \label{constant}
\end{center}
\vspace{-0.5cm}
\end{figure}

The general location of the SUSY vacuum as a function of the parameter can be calculated analytically by solving $D_\Phi W = 0$. Requiring this solution to be located within the boundaries of moduli space puts the following constraints on the parameter:
 \begin{align}
   c \in (1.56403, 1.93276) \,.
 \end{align}
Moreover, the value of the CC, inside this SUSY vacuum, can also be calculated, leading to a parabola centered around $c = \sqrt{3}$. For values close to the maximum it can be approximated by
 \begin{align}
   V_0 = -3 m^2 (1 + (c-\sqrt{3})^2 - \tfrac89 \sqrt{3} (c - \sqrt{3})^3 + \ldots ) \,.
 \end{align}
However, this only yields the CC for values of $c$ within the above range. For other values, there is only a non-SUSY vacuum within the fundamental strip. The CC then behaves as is plotted in Fig.~\ref{constant}.
 
For sufficiently large deviations from the central value $c= \sqrt{3}$, one can achieve uplifting to De Sitter values, passing through Minkowski. This happens at the points 
\begin{align}
{\rm Mink:~~} c=0 \,, \qquad c = 3.67044 \,.
\end{align}
Can either of these be used to model a dark energy vacuum that one ends up in after a successful inflationary stage? A first check to this end is the value of the inflationary energy. Setting $\phi=0$, the potential takes the value (included as a light yellow line in the figure)
  \begin{align}
   V = c ( c - 2 \sqrt{3}) \,.
  \end{align}
Note that it intersects the CC in two locations: the non-SUSY Minkowski for $c=0$ and the SUSY AdS for $c=\sqrt{3}$. Indeed these are exactly the parameter values for which the minimum is located at $\phi=0$ and hence on the vertical axis. This implies that the Minkowski vacuum with $c=0$ cannot be employed for inflation (this indeed corresponds to the first case described in Sec.~\ref{Wlin}). In contrast, the other Minkowski vacuum is located at a position when the inflationary energy has a non-zero, positive value. Thus, there are parameter values close to $c = 3.67044$ that allow for an arbitrarily small CC in the vacuum, while the inflationary energy can be tuned independently.

\end{document}